# A traffic model based on fuzzy cellular automata

BARTŁOMIEJ PŁACZEK

placzek.bartlomiej@gmail.com

*Institute of Computer Science, University of Silesia,*

*Będzińska 39, 41-200 Sosnowiec, Poland*

Cellular automata (CA) play an important role in the development of computationally efficient microscopic traffic models and recently have gained considerable importance as a mean of optimising traffic control strategies. However, real-time application of the available CA models in traffic control systems is a difficult task due to their discrete and stochastic nature. This paper introduces a novel method for simulation of signalised traffic streams, which combines CA and fuzzy numbers. The introduced traffic simulation algorithm eliminates main drawbacks of the CA approach, i.e. necessity of multiple Monte Carlo simulations and calibration issues. Computational cost of traffic simulation for the proposed algorithm is considerably lower than the cost of simulation based on stochastic CA. Thus, the simulation results can be obtained in a much shorter time. Experiments confirmed that the simulation results for the introduced algorithm are consistent with that observed for stochastic CA. The proposed simulation algorithm is suitable for real-time applications in traffic control systems.[*]

## 1 INTRODUCTION

Cellular automata (CA) have become a frequently used tool for microscopic modelling of road traffic processes. However, the development of CA-based traffic simulation algorithms for applications in road traffic control is still a challenging issue. Such algorithms should present a well-balanced trade-off between accuracy and computational complexity to enable on-line processing of measurement data and traffic state estimation.

The state-of-the-art traffic control methods use macroscopic and mesoscopic models that describe queues or groups of vehicles [1, 2, 3]. Unfortunately, these models ignore the individual vehicle parameters (position, velocity, acceleration, direction, etc.) that are important from the traffic control point of view, as they have a significant influence on the traffic performance.

Traffic data concerning the parameters of particular vehicles are available in modern sensing platforms (e.g. vision-based monitoring systems [4, 5] and vehicular sensor networks [6, 7]). Emerging technologies in the road traffic monitoring enable wireless communication between sensing devices installed in vehicles (mobile sensors) and the road environment for dynamic transfers of measurement data [8]. Thus, the vehicles can be used as sources of information to determine detailed traffic stream characteristics. These data cannot be fully utilised for traffic control purposes when using the macroscopic or mesoscopic models.

Sufficiently accurate microscopic traffic models have been developed based on the CA theory [9, 10]. However, application of the available CA models in traffic control systems is a difficult task due to their stochastic character. Stochastic parameters in the CA are necessary to represent model uncertainties and to enable model calibration. Unfortunately, the traffic simulation with stochastic CA requires the time consuming Monte Carlo method to be used. Long computational time of the Monte Carlo simulation is a critical disadvantage for the traffic control applications that require the results of simulation to be obtained in strictly limited time-frames.

In this paper a new CA-based simulation algorithm is introduced, which does not involve the Monte Carlo technique and enables a realistic simulation of signal controlled traffic streams. The simulation algorithm was formulated based on the CA approach and fuzzy logic. The original feature distinguishing this algorithm from state-of-the-art methods is that vehicle positions, velocities and other parameters are represented by fuzzy numbers. Moreover, two different deterministic rules of CA are used simultaneously to update both the fuzzy velocities and fuzzy positions.

The application of fuzzy numbers helps to deal with imprecise traffic data and to describe uncertainty of the simulation results. In fact, it is impossible to predict unambiguously the evolution of a traffic stream. Moreover, the current traffic state also cannot be usually identified precisely on the basis of the available measurement data. Therefore, the fuzzy numbers are used in the traffic simulation algorithm to describe the uncertainty and imprecision of the simulation inputs and outputs. The algorithm allows a single simulation to take into account many potential scenarios (traffic state evolutions) [11]. These facts along with a low computational complexity make the algorithm suitable for real-time applications in traffic control systems.

The rest of the paper is organised as follows: related works are reviewed and analysed in Section 2. Section 3 addresses limitations regarding the use of CA for modelling the traffic at signalised intersections that have motivated the research reported in this paper. Section 4 introduces the new CA-based algorithm for simulation of signalised traffic stream. Section 5 includes a comparison of computational costs as well as simulation results for the proposed algorithm and the Nagel-Schreckenberg stochastic CA. Finally, conclusions are given in Section 6.

## 2 RELATED WORKS

Among many works on CA applications in the field of road traffic modelling [12, 13], there are also studies that use CA for simulation and optimisation of a signal traffic control. In [14] a traffic simulation tool for urban road networks was proposed which is based on the Nagel-Schreckenberg (NaSch) stochastic CA [15]. An intersection model was considered in this work including traffic regulations (priority rules, signs, and signalisation). It was also suggested that for appropriate setting of a deceleration probability parameter the model yields realistic time headways between vehicles crossing a stop line at a signalised intersection.

A modified NaSch model for a traffic flow controlled by traffic signals was proposed in [16]. According to the introduced modification, the deceleration probability for each vehicle is determined as a function of free space in front of the vehicle. The model was applied in order to simulate a signal controlled traffic flow on a single-lane road. Several models of this type that are based on the NaSch CA can be found in the literature (e.g.: model with turning-deceleration rule [17], model with anticipation of change in traffic lights [18]).

Schadschneider et al. [19] have presented a CA model of a vehicular traffic in signalised urban networks by combining ideas borrowed from the Biham-Middleton-Levine model of city traffic [20] and the NaSch model of a single lane traffic stream. A similar model was adopted to calculate optimal parameters of a traffic signal coordination plan that maximise the flow in a road network [21].

The stochastic CA models applied in traffic simulation tools were further extended to agent-based models that aim at reproducing sophisticated driver behaviours [22]. In this approach the drivers are represented by autonomous agents able to make complex decisions about route planning.

In [23] a model of city traffic was introduced, which is based on deterministic elementary CA. The simplicity of this model allows for the simulation of large road networks with many intersections. Another method of road network modelling uses rings of cells as a simplified representation of intersections [24]. The emphasis in those approaches was put on the simplicity and scalability of the model rather than on realism of the traffic simulation. A calibration method discussed in [25] allows the computationally simple CA models to reproduce realistic traffic dynamics on a macroscopic level.

CA traffic model was also utilised as an evaluation tool in a genetic algorithm for the traffic signals optimisation [10]. The fitness function in this algorithm is evaluated on the basis of traffic simulation results. The optimisation was performed for a road network with 20 signalised intersections. The results were compared for a stochastic and a deterministic version of the CA model. It was observed that the obtained population fitness ranking is similar for both versions. However, the deterministic CA have enabled a remarkable speed-up of the genetic algorithm execution.

The relationships between parameters of the CA models and saturation flow rates at a simulated intersection were analysed in [26]. The traffic models were investigated in this work to identify the possibility of reproducing any desired value of the saturation flow. This analysis was performed for both the deterministic and stochastic CA. It was concluded that the stochastic version allows any value of the saturation flow to be obtained by adjusting a deceleration probability parameter.

Main drawbacks of the above CA models, that impede their practical applications in road traffic control, are discussed in the next section.

The proposed method combines a CA model of road traffic with fuzzy logic. Hybrid systems that combine CA and fuzzy sets are typically referred to as fuzzy cellular automata (FCA) [27]. FCA-based models have found many applications in the field of complex systems simulation [28, 29]. A road traffic model of this kind has been proposed in [30]. In such models, the local update rule of a classical CA is usually replaced by a fuzzy logic system consisting of fuzzy rules, fuzzification, inference, and defuzzification mechanisms. A different approach is used in this paper: current states of the cells are determined by fuzzy numbers and two deterministic CA rules are involved in the update operation. The innovative features of the proposed method are the elimination of information loss caused by defuzzification and the incorporation of uncertainty in simulation results.

# 3 LIMITATIONS OF CA TRAFFIC MODELS

CA models of road traffic describe velocities and positions of vehicles in discrete time steps. Position $x_{i,t}$ indicates a cell, which is occupied by vehicle $i$ at time step $t$. Velocity $v_{i,t}$ is expressed in cells per time step and determines how many cells the vehicle $i$ advance at time step $t$ ($v_{i,t} = x_{i,t+1} - x_{i,t}$). The discrete positions and velocities are updated at each time step according to a rule of CA. In order to compute velocity, the rule takes into account previous velocity values, a maximal velocity $v_{max}$ and the number of free cells in front of the vehicle (so-called gap) $g_{i,t}$. Let $c_{k,t}$ denote the state of $k$-th cell in a CA traffic model at time step $t$. For empty cells it is assumed that $c_{k,t} = -1$. For occupied cells the cell state corresponds to the vehicle's velocity, i.e. $c_{k,t} = v_{i,t} \Leftrightarrow x_{i,t} = k$.

Realistic simulation of signal controlled traffic streams requires a traffic model, which can be appropriately calibrated to reflect real values of *saturation flow rate*, i.e. the number of vehicles that would pass through a signalised intersection during a given time period (usually one hour), if the traffic signal stayed green for the entire period. For CA traffic models the calibration is not a trivial task due to their discrete formulation and a limited set of parameters.

## 3.1 Deterministic cellular automata

In case of deterministic CA traffic models the saturation flow rate is constant and it can be evaluated by an analysis of queue discharge behaviour. To this end a traffic stream have to be modelled, which consists of vehicles leaving a queue after end of red signal. In such traffic stream a uniform gap $g$ exists between vehicles that reached a constant maximal velocity $v_{\max}$. Every vehicle occupies one cell, thus the gap $g$ corresponds to the traffic density of $1/(g + 1)$ vehicles per cell. On this basis the saturation flow rate $s$ can be calculated in vehicles per time step [26]:

$$s = \frac{v_{\max}}{g+1}. \tag{1}$$

According to the above equation, the saturation flow in a deterministic CA model can be adjusted in two ways: by changing the maximal velocity or changing the gap that occurs between vehicles moving with the maximal velocity. Such changes require a modification of the CA rule. In practice, it is not possible to obtain an arbitrary saturation flow rate because $v_{max}$ and $g$ takes only integer values and a set of applicable rules that realistically reproduce the traffic behaviour is very limited.

In order to illustrate the issue of deterministic CA traffic model calibration, the saturation flow rate will be analysed taking into account two different rules (R1, R2) that use following formulas to calculate the positions and velocities of vehicles:

$$x_{i,t+1} = x_{i,t} + v_{i,t}, \tag{2}$$

$$v_{i,t} = \begin{cases} u_{j,k}, & u_{j,k} \geq 0 \\ g_{i,t-1} & u_{j,k} = -1 \end{cases}$$

where: $u_{j,k}$ is an element of a matrix $U$ with 4 rows and 6 columns, $j = v_{i,t-1} + 1$, and $k = \min(g_{i,t}, 5) + 1$. Let $U^{\mathrm{I}}$ and $U^{\mathrm{II}}$ denote the matrices defined for rules R1 and R2 respectively:

$$U^{\mathrm{I}} = \begin{bmatrix} 0 & -1 & 1 & 1 & 1 & 1 \\ 0 & 1 & 1 & 1 & 2 & 2 \\ 0 & 1 & 1 & 1 & 2 & 2 \\ 0 & 1 & 1 & 1 & 2 & 2 \end{bmatrix}, \quad U^{\mathrm{II}} = \begin{bmatrix} 0 & -1 & 1 & 2 & 1 & 1 \\ 0 & 1 & 1 & 2 & 2 & 2 \\ 0 & 1 & 1 & 2 & 3 & 2 \\ 0 & 1 & 1 & 2 & 3 & 2 \end{bmatrix}. \tag{3}$$

For instance, if the velocity of a vehicle at previous time step ($v_{i,t-1}$) was 1 and the gap $g_{i,t}$ is 3 then the current velocity of such vehicle is determined by $u_{2,4}$. It means that the vehicle will advance by 1 cell when using rule R1 and by 2 cells when using rule R2. The analysed rules assume that a stopped vehicle, which has only one empty cell in front of it, moves forward with delay of one time step. Thus, the velocity in this case has to be computed in a different way and the value –1 was used in matrices (3) to distinguish this situation.

Fig. 1 compares the queue discharge behaviour for the considered rules. The numbers in cells denote states of the occupied cells, i.e. velocities of vehicles. Symbol “X” represents a red signal for vehicles, which is active only at the first time step of the simulation. When applying the first rule (R1), the resulting gap $g$ between vehicles in saturated traffic stream is of 4 cells (Fig. 1 a). A smaller gap ($g$ = 3 cells) is obtained for rule R2 (Fig. 1 b). In both cases, the maximal velocity is 2 cells per time step. The resulting saturation flow rate $s$ equals 0.4 [vehs/time step] = 1440 [vehs/h] for rule R1 and 0.5 [vehs/time step] = 1800 [vehs/h] for rule R2. The saturation flow rates in vehicles per hour of green time were calculated assuming that one time step corresponds to one second.

The presented examples show that it is possible to obtain only a limited set of the saturation flow rates by manipulating rules of the deterministic CA models. It should be noted that the minimal modification of model parameters ($g$ was increased by 1, $v_{max}$ was constant) results in a significant change of the saturation flow rate (*s* increased by 360 [vehs/h]). Thus, the deterministic CA models are not sufficient for the realistic traffic simulation at signalised intersections.

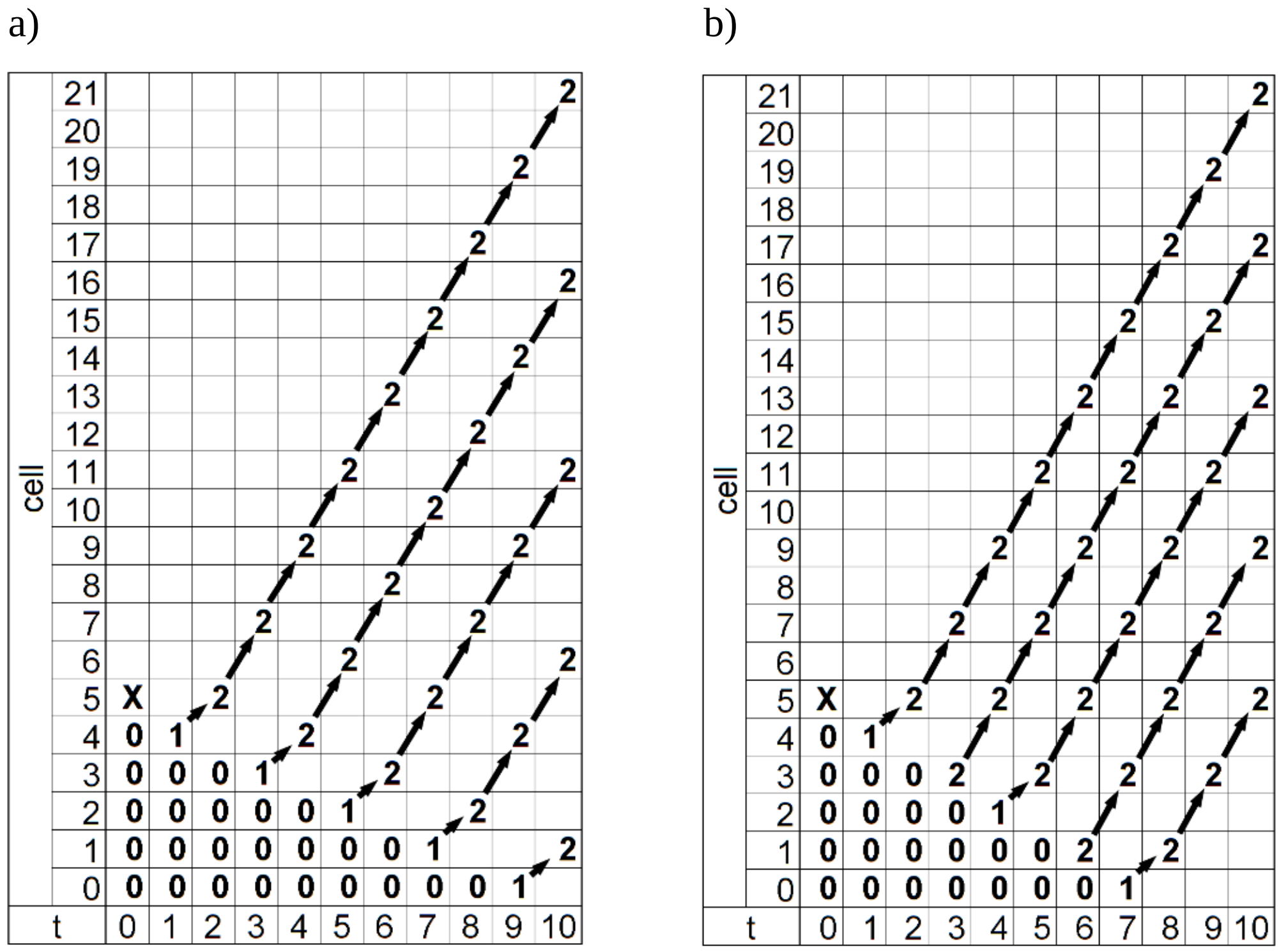

Fig. 1. Queue discharge behaviour for deterministic CA rules R1 (a) and R2 (b)

### 3.2 Stochastic cellular automata

In stochastic CA traffic models, the rule includes an additional probability parameter *p*, which controls random aspects of the vehicles acceleration (so-called deceleration probability).

When considering the real-time applications, an important drawback of the stochastic CA models is the necessity of using the Monte Carlo method [10]. This relates to the fact that in practice a number of traffic simulation runs is necessary to obtain significant and useful results. Therefore, the applicability of the stochastic CA is limited due to the long computational time of the Monte Carlo simulation. This disadvantage is critical in the traffic control applications, where the simulation results have to be obtained much faster than the real duration time of the simulated process.

Calibration of a traffic model based on the stochastic CA may also be difficult. As it was discussed in [26], the probability parameter *p* has a direct influence on saturation flow rates of the simulated traffic stream. However, modification of *p* results not only in change of the average (expected) saturation flow rate, but also in change of its spread. This effect is illustrated by experimental results obtained for the NaSch model (Figs. 2 and 3). During the experiment, the probability parameter *p* was changed between 0 and 0.8 with increments of 0.01. The simulation of a one hour period was repeated five hundred times for every value of the probability *p*. Saturation flow rate was calculated in each simulation run. On this basis a distribution of the saturation flow rates was determined for every value of parameter *p* in the analysed range. The plot in Fig. 2 shows medians, 5-th, and 95-th percentiles of the obtained saturation flow distributions. An example of the distribution histogram for $p = 0.2$ is presented in Fig. 7. The spread of the saturation flow rates was evaluated as a difference between 95-th and 5-th percentile (Fig. 3).

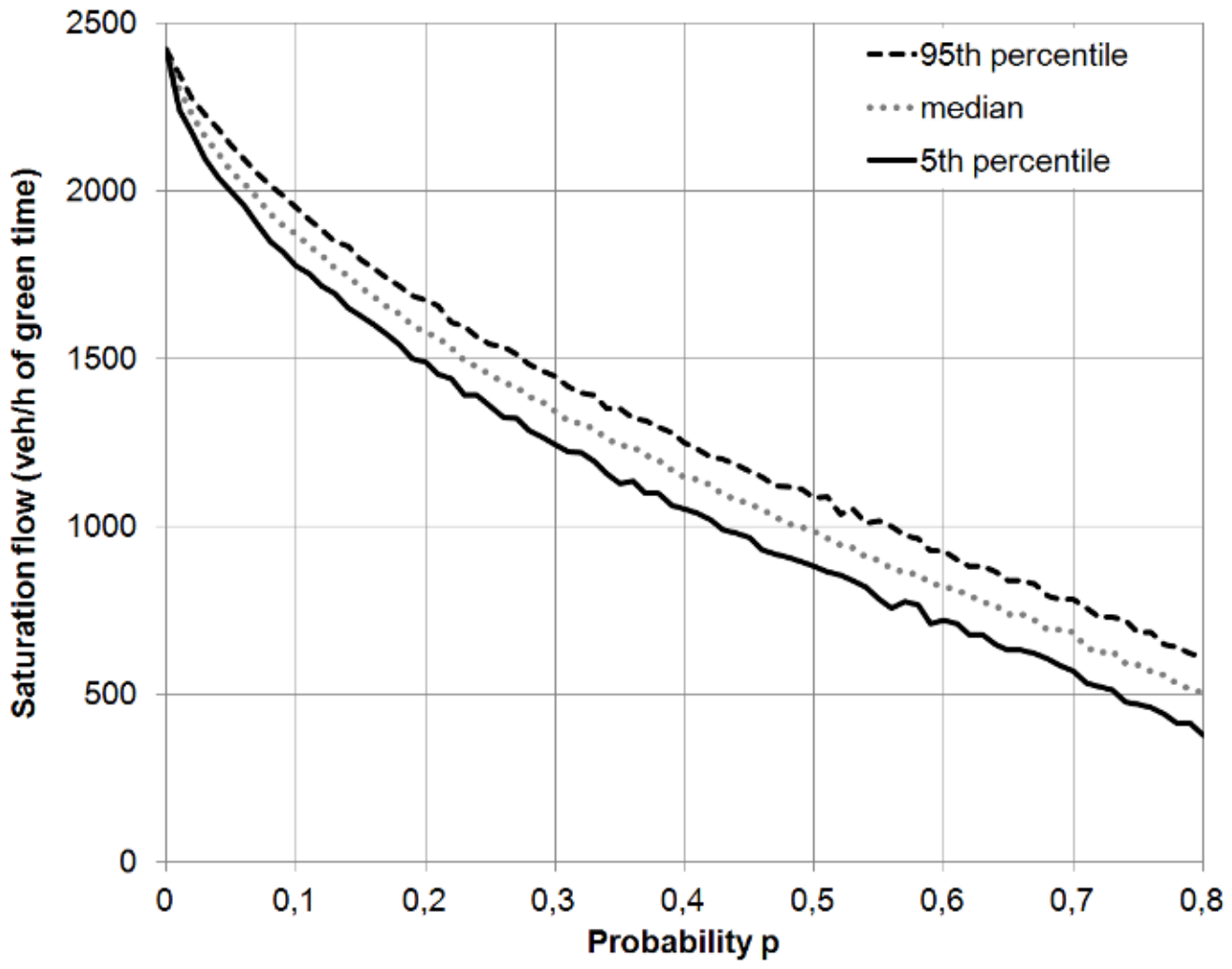

Fig. 2. Saturation flow rate vs. probability parameter *p* for NaSch model

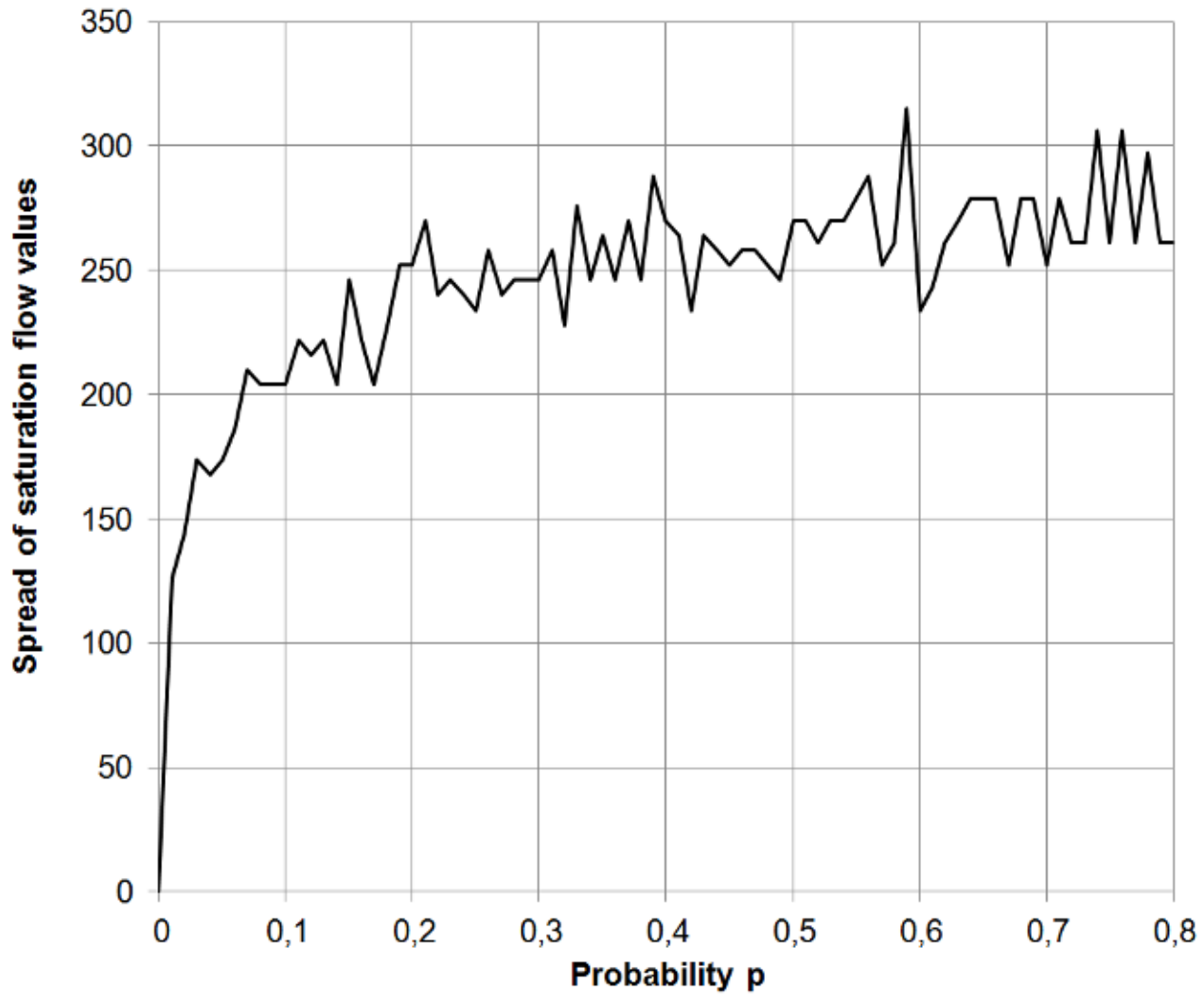

Fig. 3. Spread of saturation flow values vs. probability parameter *p* for NaSch model

From the results in Figs. 2 and 3 it may be concluded that an increase of the probability $p$ causes both lower saturation flow rates and higher spread of its values. Thus, for the NaSch model it is impossible to independently change the average value and the range of the saturation flow rates.

Another issue is related to the dependency between free-flow velocity $v_f$ and the probability parameter: $v_f = v_{\max} - p$, which is a well-known characteristic of the NaSch CA [12]. Due to this dependency, any modification of the probability parameter results also in a change of the free-flow velocity. The aforementioned mutual dependencies between parameters in the stochastic CA seriously impede the use of the probability parameter for model calibration.

## 4 TRAFFIC SIMULATION ALGORITHM

A new CA-based traffic simulation algorithm was developed to overcome the limitations of CA models that were discussed in previous section. The introduced algorithm has a low computational complexity and allows parameters of the simulated traffic stream to be precisely calibrated.

The most important components of the algorithm are update rules that enable the velocities and positions of vehicles to be computed at each time step of the simulation. Stochastic CA rules are not applicable here as one of the objectives is to eliminate the necessity of Monte Carlo simulations. The proposed traffic simulation algorithm utilises a pair of deterministic rules (RL, RH) for updating the CA model. Rules RL and RH correspond to two extreme saturation flow values. Using the introduced simulation method an arbitrary saturation flow rate can be obtained, which lies in between these two extremes.

The rules RL and RH are used to compute auxiliary velocities and positions of vehicles, denoted by $v_{i,t}^{L}, x_{i,t+1}^{L}$ for rule RL and $v_{i,t}^{H}$, $x_{i,t+1}^{H}$ for rule RH, where $t$ is current time step of the simulation. As it was discussed in Section 3.1, each deterministic CA rule corresponds to a single value of the saturation flow. The value of saturation flow depends on two parameters: maximal velocity $v_{\max}$ and minimal gap $g$ between vehicles that move with the maximal velocity. We will denote the maximal velocity for rules RL and RH by $v_{\max}^{L}$ and $v_{\max}^{H}$ respectively. Similarly, the gaps for rules RL and RH will be represented by $g^L$ and $g^H$. The corresponding saturation flow values can be calculated by using Eq. (1). In the proposed algorithm the saturation flow level $s^L$, achieved for rule RL, has to be lower than the saturation flow level $s^H$ for the second rule (RH).

Resultant velocity $v_{i,t}$ and position $x_{i,t+1}$ of a vehicle are calculated by using one of the available rules (RL or RH). In order to decide which rule has to be selected at a given time step for a particular vehicle, the proposed algorithm takes into account a normalised position:

$$\bar{x}_{i,t} = \frac{x_{i,t} - x_{i,t}^{L}}{x_{i,t}^{H} - x_{i,t}^{L}}. \quad (4)$$

Note that $\bar{x}_{i,t} \in [0, 1]$ since the resultant position is always bounded by the auxiliary positions: $x_{i,t}^{L} \le x_{i,t} \le x_{i,t}^{H}$.

The objective of rule selection is to obtain a constant normalised position for all vehicles at each time step of the simulation. Formally, this objective can be defined as follows:

$$\left| \bar{x}_{i,t} - \alpha \right| \to \min \ \ \forall i \ \ \forall t, \quad (5)$$

where $\alpha \in [0, 1]$ is a calibration parameter of the simulation algorithm, which corresponds to an expected value of the normalised position. During traffic simulation it is usually impossible to obtain $\bar{x}_{i,t} = \alpha$ because the positions $x_{i,t}$ have to be computed as integer cell indices in accordance with the deterministic CA rules. Exact definition of the rule selection operation is presented in the form of if-then statement by the pseudo-code of the proposed simulation algorithm (Algorithm 1).

The calibration parameter $\alpha$ is used to adjust the saturation flow rate of the simulated traffic stream. Thus, the dependency between parameter $\alpha$ and the saturation flow rate has to be known. In order o determine this dependency, we will assume that for a saturated traffic stream the normalised vehicle position is constant, i.e. $\bar{x}_{i,t} = \alpha \ \ \forall i \ \ \forall t$. Therefore the resultant position of a vehicle can be expressed as:

$$x_{i,t} = x_{i,t}^{L} + \alpha (x_{i,t}^{H} - x_{i,t}^{L}) \ \ \forall i \ \ \forall t. \quad (6)$$

Since all the vehicles move with velocity $v_{\max}$, the following equality is satisfied:

$$v_{\max} = x_{i,t} - x_{i,t-1}. \quad (7)$$

By substituting (6) into (7), after simple transformations, we have:

$$v_{\max} = v_{\max}^{L} + \alpha (v_{\max}^{H} - v_{\max}^{L}). \quad (8)$$

As it was mentioned above, for deterministic CA all gaps between vehicles in a saturated traffic stream have the same length, which can be calculated using the formula:

$$g = x_{i,t} - x_{i+1,t} - 1. \quad (9)$$

From Eqns. (6) and (9) we obtain

$$g = g^L + \alpha(g^H - g^L)\,. \tag{10}$$

Finally, by substituting (8) and (10) into (1) we can express the saturation flow rate as a function of the calibration parameter $\alpha$:

$$s = \frac{v_{\max}^L + \alpha(v_{\max}^H - v_{\max}^L)}{g^L + \alpha(g^H - g^L) + 1}\,. \tag{11}$$

Thus, in order to obtain a predetermined saturation flow rate $s$ for the simulated traffic stream the calibration parameter $\alpha$ must be set according to the following formula:

$$\alpha = \frac{s(g^L + 1) - v_{\max}^L}{(v_{\max}^H - v_{\max}^L) - s(g^H - g^L)}\,. \tag{12}$$

For realistic traffic simulation it is necessary to take into account a random acceleration of vehicles, which results in a random level of the saturation flow. In case of the stochastic CA, this randomness is represented by the probabilistic update rule with the deceleration probability parameter. To eliminate the necessity of using a stochastic CA rule, the introduced simulation algorithm describes the uncertain positions and velocities of vehicles by means of fuzzy numbers. In consequence, the saturation flow rate of the simulated traffic stream is also expressed by a fuzzy number. This approach allows the traffic model to be calibrated in order to reflect real values and uncertainties of measured saturation flows. Advantages of this approach were discussed in previous works [11, 31].

Without loss of generality, it was assumed that triangular fuzzy numbers will be used in this study for presentation and evaluation of the simulation algorithm. The triangular fuzzy numbers are represented by three components. Fig. 4 shows a membership function of a fuzzy number $Z = (z^{(1)}, z^{(2)}, z^{(3)})$.

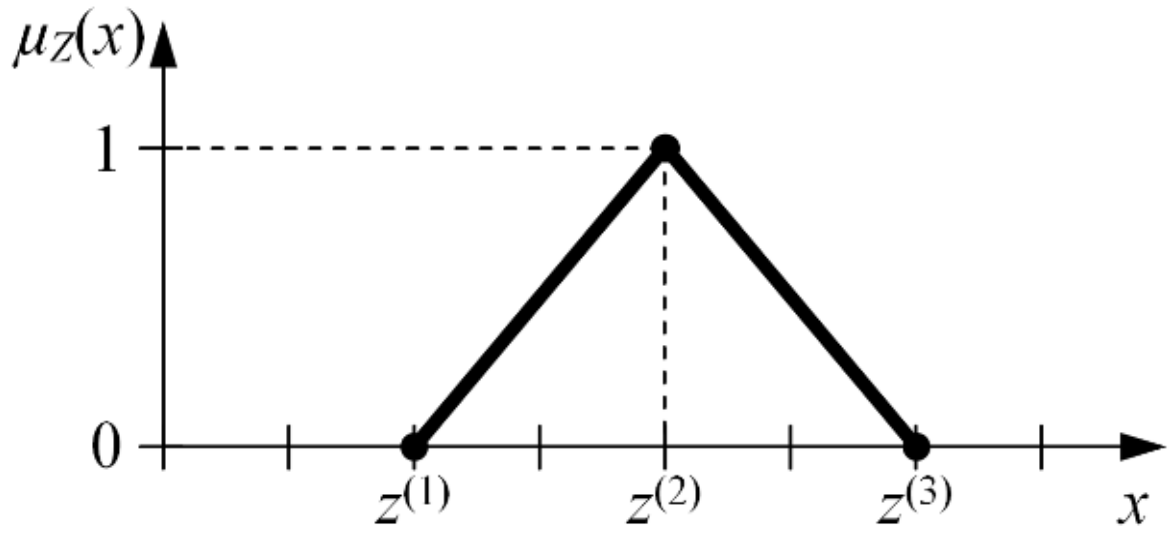

Fig. 4. Triangular fuzzy number

The aim of the introduced approach is to provide a road traffic simulation algorithm, which can accept a fuzzy number $S = (s^{(1)}, s^{(2)}, s^{(3)})$ as an input parameter specifying the level of the saturation flow. Therefore, the method discussed above was extended to take into account the three components that are necessary for representation of the triangular fuzzy numbers. According to this modification, velocities and positions of vehicles are described by fuzzy numbers.

Update of fuzzy velocity $V_{i,t} = (v_{i,t}^{(1)}, v_{i,t}^{(2)}, v_{i,t}^{(3)})$, and fuzzy position $X_{i,t+1} = (x_{i,t+1}^{(1)}, x_{i,t+1}^{(2)}, x_{i,t+1}^{(3)})$ is executed by using the deterministic rules RL and RH. The rule is selected independently for each component of the fuzzy number representation. Objective of the selection operation is consistent with that defined by (5). One difference is that the operation is executed three times with respect to the components of normalised position $\bar{x}_{i,t}^{(m)}$ and the calibration parameters $\alpha^{(m)}$ ($m$ = 1, 2, 3). The calibration parameters $\alpha^{(m)}$ have to be calculated according to Eq. (12) with taking into account the three components of the assumed fuzzy saturation flow rate *S*.

The new CA-based simulation algorithm described above is summarized in the following pseudocode:

Algorithm 1. Simulation of signal controlled traffic stream

```
For t = 1 to T do
  Update traffic signals.
  For all vehicles (i = 1 to N) do
    Compute v^L_{i,t} and x^L_{i,t+1} using rule RL.
    Compute v^H_{i,t} and x^H_{i,t+1} using rule RH.
    For m = 1 to 3 do
      if x̄^(m)_{i,t} ≤ α^(m) then compute v^(m)_{i,t} and x^(m)_{i,t+1} using rule RH,
      else compute v^(m)_{i,t} and x^(m)_{i,t+1} using rule RL.
```

Fig. 5 illustrates the concepts that underlie the proposed algorithm and concern the determination of vehicles positions in a saturated traffic stream. Upper chart in Fig. 5 shows non-linear relationships between positions *x* of particular vehicles moving in a traffic stream and the saturation flow *s* that correspond to Eq. (11). The lower chart in Fig. 5 shows positions of vehicles that were obtained by taking into account the assumed fuzzy value of the saturation flow *S*. These positions are represented by fuzzy numbers $X_i$. Black dots represent two extreme configurations of the traffic model that are obtained by using the two different deterministic CA rules (RL and RH).

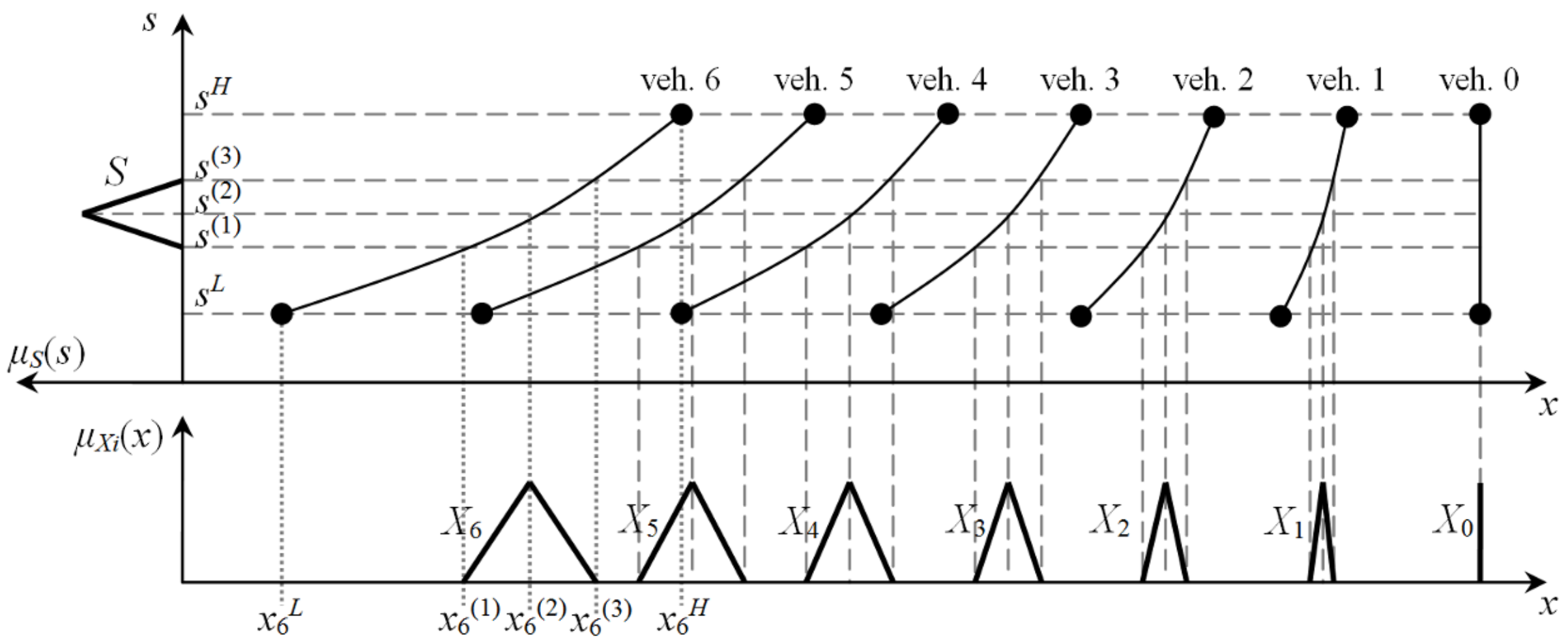

Fig. 5. Fuzzy positions of vehicles in saturated traffic stream

Since the presented algorithm is designed to be used for the simulation of signal-controlled traffic streams, it has to take into account drivers' reactions to traffic signals. The influence of traffic signalisation is simulated by introducing a set of cells in front of which the vehicles have to halt at red signals. The halt cells correspond with the locations of stop lines. The update of a traffic signal involves the insertion of an appropriate cell index $x$ into the set of halt cells when the red signal has to be activated, and removal of the index $x$ from that set when the red signal has to be deactivated. Thus, the yellow time is considered as a part of the green phase.

# 5 COMPARISON WITH STOCHASTIC CA MODEL

In this section the proposed simulation algorithm is compared against an algorithm, which was based on the NaSch stochastic CA. The comparison takes into account computational costs as well as results of traffic simulation for a signalised arterial.

## 5.1 Computational cost

For comparison purpose, the stochastic rule of NaSch CA [15] was decomposed into two deterministic rules, denoted by NSH and NSL. The NSH rule corresponds to the original NaSch rule with deceleration probability $p = 0$. According to this rule, velocity of a vehicle is calculated using the following formula:

$$v_{i,t} = \min(v_{i,t-1} + 1, g_{i,t}, v_{\max}) . \quad (13)$$

The NSL rule reflects the operation of the NaSch rule for $p = 1$. Thus, the velocity in this rule is decreased by 1:

$$v_{i,t} = \max(0, \min(v_{i,t-1} + 1, g_{i,t}, v_{\max}) - 1) . \quad (14)$$

The randomisation step of the NaSch model was implemented in the simulation algorithm by introducing a selection of the deterministic rule (NSL or NSH). The selection is based on a random number $\xi \in [0,1)$, which is drawn from a uniform distribution. Such description of the NaSch model (Algorithm 2) enables its comparison with the proposed simulation algorithm (Algorithm 1).

Algorithm 2. Traffic simulation with stochastic NaSch CA

```
For simulation run 1 to K do
  For t = 1 to T do
    Update traffic signals.
    For all vehicles (i = 1 to N) do
      Generate random number ξ.
      If ξ < p then compute v_{i,t} and x_{i,t+1} using rule NSL,
      else compute v_{i,t} and x_{i,t+1} using rule NSH.
    Store simulation results.
```

The implementation of the NaSch model (Algorithm 2) requires multiple traffic simulation runs. At each run, the simulation results have to be stored. After $K$ runs, the stored results are used to calculate distributions of the traffic performance measures. The number of simulation runs $K$ has to be appropriately high in order to obtain meaningful estimates. For the experiments presented in this study the number of runs $K$ was 500.

Let us assume that the basic operation in the traffic simulation algorithm is the execution of a deterministic CA rule, i.e. the computation of the position and velocity for a single vehicle. The number of basic operations performed during the traffic simulation can be determined for both compared models by analysing the pseudo-code of Algorithm 1 and Algorithm 2. The traffic simulation with the NaSch model requires $K \cdot T \cdot N$ basic operations whereas during the simulation with the proposed algorithm the basic operation is executed $5 \cdot T \cdot N$ times. The factor 5 relates to the fact that the CA rule has to be used 5 times to determine the fuzzy velocity and fuzzy position of each vehicle (see Algorithm 1). It was assumed that the number of vehicles $N$ is constant in the analysed simulation period.

The complexity of both algorithms is $O(N)$, however the computational cost of traffic simulation is considerably reduced for the proposed algorithm (Algorithm 1) because the number of simulation runs $K$ is always much greater than 5 (usually amounts to several hundred runs). Moreover, the proposed simulation algorithm does not need to store partial results, thus it requires less memory space than the simulation with the NaSch CA.

## 5.2 Simulation results

This section presents some results of traffic simulations in a signalised arterial. The results of a simulation performed with the proposed algorithm were compared against those obtained by using the stochastic NaSch CA.

For the purpose of this experiment the rules denoted by RL and RH in the proposed simulation algorithm (Algorithm 1) are implemented by using the definition of rules R1 and R2 from Sect. 3.1. It should be noted here that rules RH and RL were designed to enable transition between the two extreme saturation flow rates ($s^L$ and $s^H$). Fig. 6 includes an example of vehicles trajectories that were obtained by using the rules RL and RH. At time step 0 the rule RL is used and the saturation flow has the low value ($s^L$). After switching to rule RH, at time step 4 the saturation flow takes the high value ($s^H$). Then, by using rule RL the saturation flow rate is lowered back to $s^L$ at time step 8. Such compatibility of the rules is necessary for the proposed algorithm.

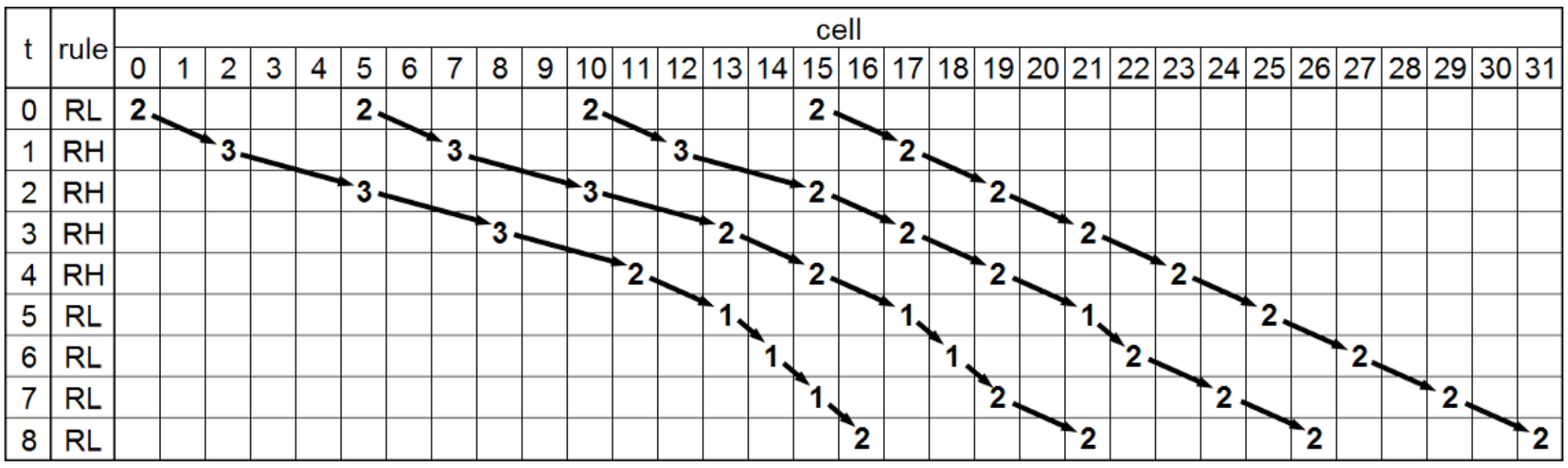

Fig. 6. Change in saturation flow rate as a result of switching CA rule

The parameters of the proposed algorithm were adjusted to match the following settings of the stochastic NaSch model: time step 1 s, cell length 7.5 m, deceleration probability $p = 0.2$, and maximal velocity $v_{\max} = 2$ cells per time step. Due to the application of rules defined in Sect. 3.1 (R1, R2) for the proposed algorithm, the maximal velocity of vehicles $V_{max}$ equals two cells per time step: $v_{max}^{(m)} = 2$, $m = 1, 2, 3$. The available interval of saturation flow values is [1440, 1800] (in vehicles per hour of green time).

The saturation flow volume in the proposed algorithm is defined by the fuzzy number $S$. This parameter was estimated in order to reproduce the distribution of saturation flow rates observed in the NaSch model. Fig. 7 shows a histogram of the saturation flow values that were obtained for the NaSch model during 500 runs of a traffic simulation at signalised intersection. The simulation period was 3600 seconds for each run. The experimental data presented in Fig. 7 were further used for the determination of the parameter $S$. The values of $s^{(1)}$, $s^{(2)}$ and $s^{(3)}$ were set respectively as the 5-th percentile, median and 95-th percentile of the saturation flow rates distribution. As a result the following fuzzy number was obtained, which describes the saturation flow in vehicles per hour of green time: $S$ = (1503, 1575, 1638). The parameters $\alpha^{(1)} = 0.21$, $\alpha^{(2)} = 0.43$, $\alpha^{(3)} = 0.60$ of the proposed algorithm were calculated for the above level of saturation flow by using the formula $\alpha = 5 - 2/s$, where $s$ is expressed in vehicles per time step. This formula was obtained from Eq. (12) and its validity was verified in simulation experiments.

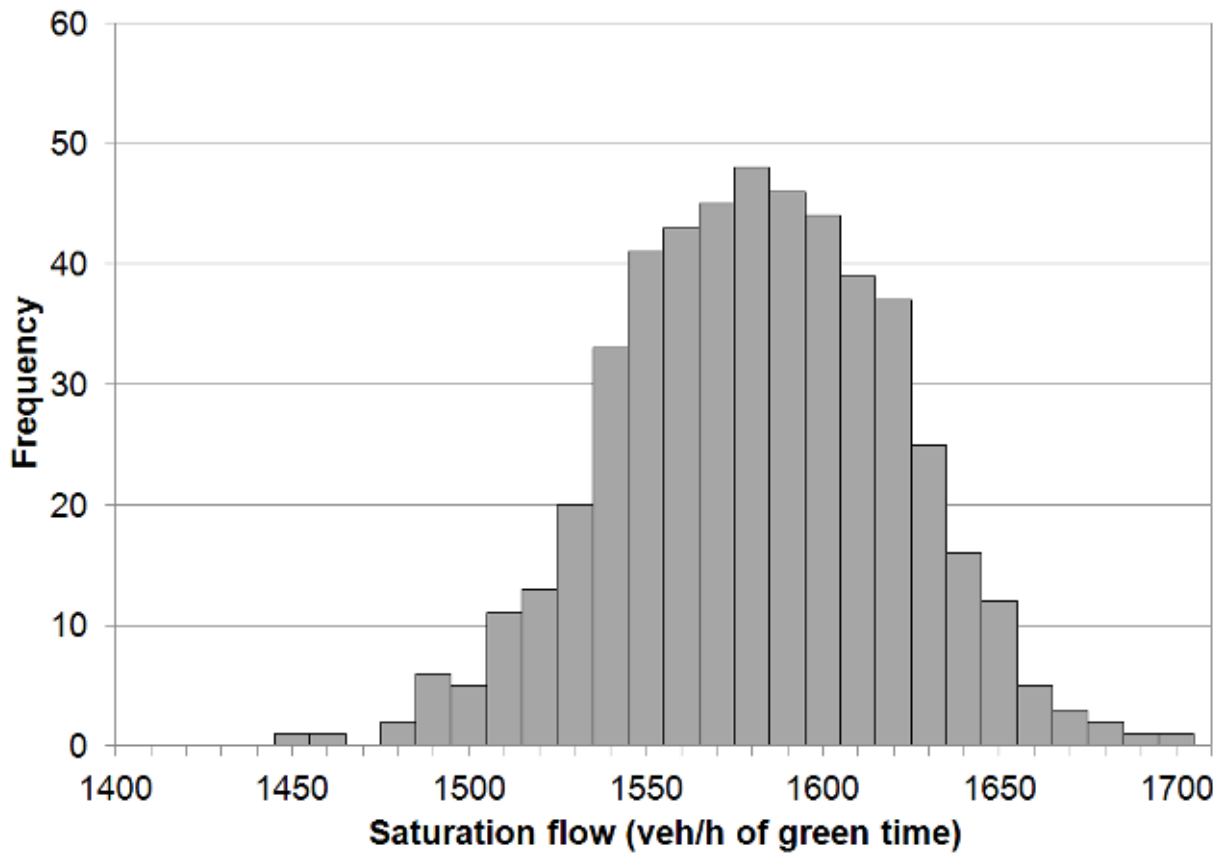

Fig. 7. Histogram of saturation flow rates for NaSch model

The cell length for the proposed algorithm was determined on the basis of free-flow velocity comparison. For low traffic densities in NaSch model, vehicles move with a free-flow velocity, which can be calculated in cells per time step using the formula: $v_f = v_{max} - p$. In case of the proposed algorithm the free-flow velocity is equal to $V_{max}$ (note that $V_{max}$ is a fuzzy number). The time step for both models corresponds to one second.

Taking into account the parameters that were determined above, the cell length for the proposed algorithm can be calculated as $7.5 \cdot (v_{\max} - p) / v_{\max}^{(m)} = 6.75$ m. Thus, in both simulation algorithms the free-flow velocity is equal to 13.5 m/s (48.6 km/h).

Simulations of traffic in a signalised one-way arterial road were performed in order to compare the introduced algorithm and the NaSch CA model. Parameters of both algorithms were set as discussed above. One-lane road was considered in this study because the analysis is focused on the detailed properties of the models that are related to single lane traffic streams. It was assumed that three signalised intersections are located at the modelled road. Total length of the road is 3 km and the distances between intersections are equal to 750 m.

The initial conditions of the traffic simulations are determined by a single queue length parameter (i.e. number of vehicles waiting in a queue at an intersection). At the beginning of each simulation queues of equal length are formed for all three intersections. Additionally, the last vehicle is always inserted into the first cell of the modelled road.

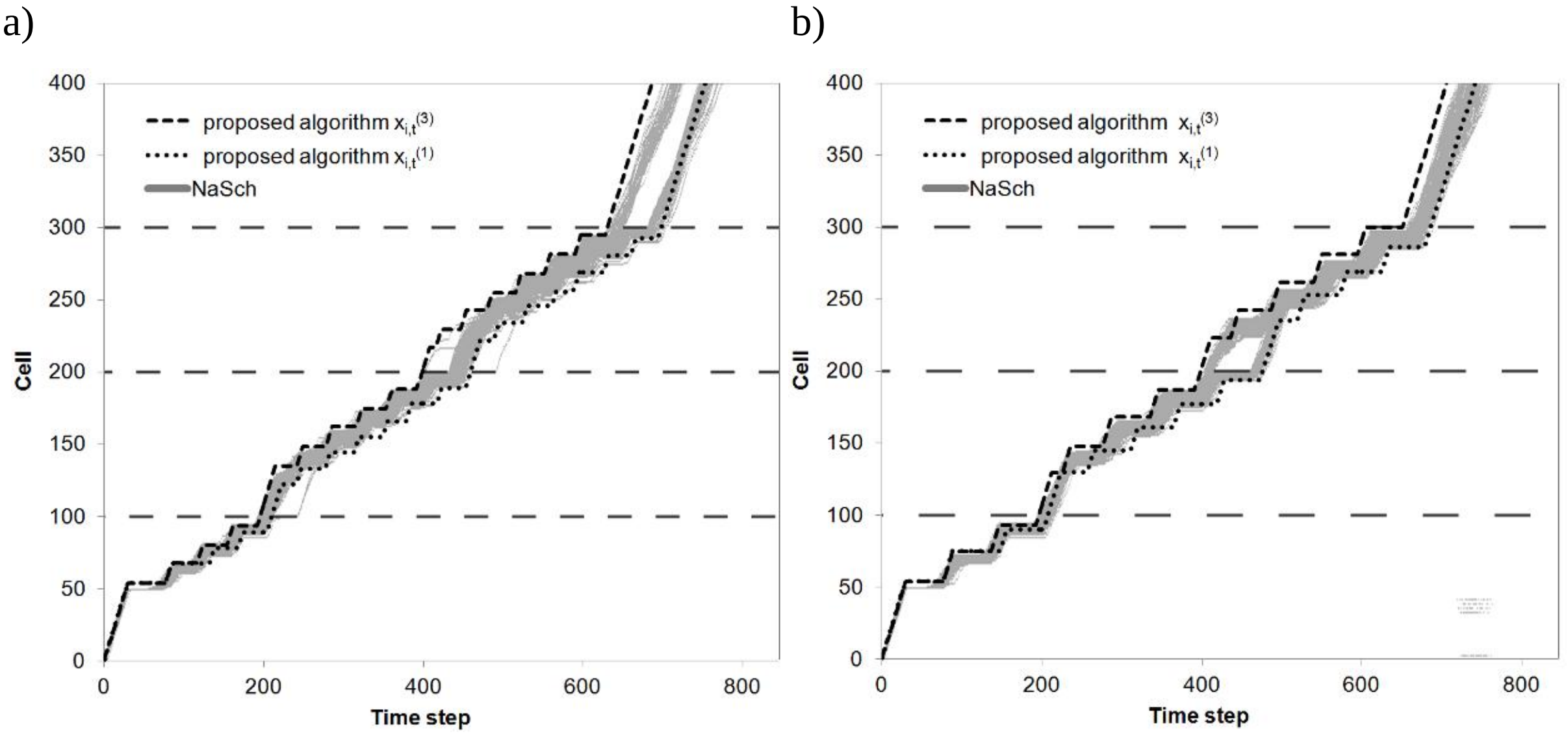

Fig. 8. Trajectory of the last vehicle for two signal cycle times: a) 60 s b) 90 s

Fig. 8 presents trajectories of the last vehicle in time-space diagrams. Black dashed and doted lines show trajectories that were determined by using the proposed algorithm. The grey colour indicates trajectories obtained for the NaSch model during 500 runs of the traffic simulation. Black horizontal bars correspond to red time intervals at the intersections. Traffic signal timings are similar for all simulated intersections. The results presented in Fig. 8 a) were obtained for signal cycle time of 60 s and green phase of 30 s. For the second case (Fig.

8 b) the cycle time was 90 s and green phase was 45 s. Offsets between the green phases at subsequent traffic signals were equal to 10 s.

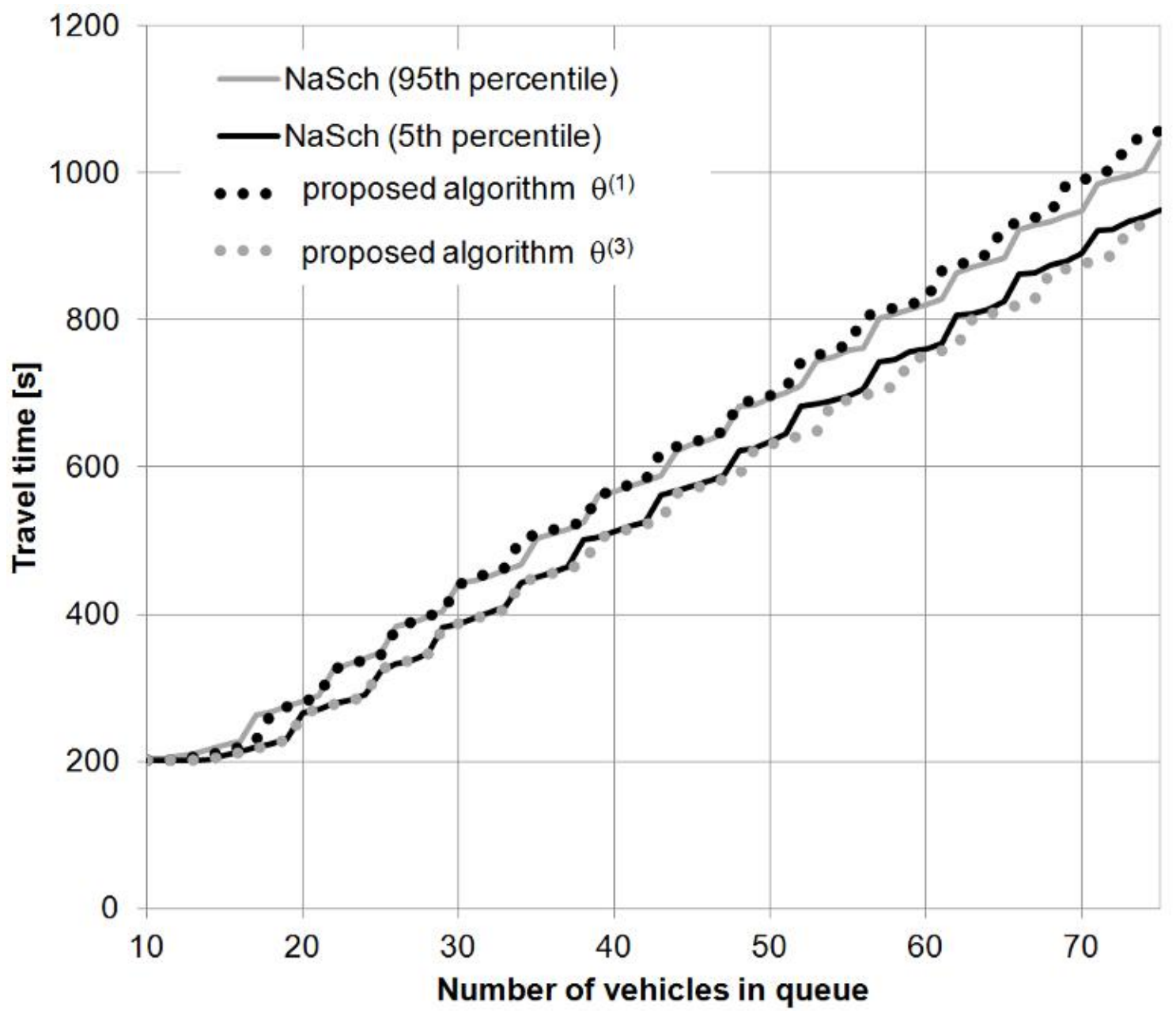

Fig. 9. Travel time of the last vehicle vs. initial queue length

Dependency between the queue length parameter and the travel time of the last vehicle is illustrated in Fig. 9. The travel time is defined as the time required for the last vehicle to pass a stop line at the third intersection. The results of travel time estimation are compared for the two analysed simulation algorithms. In case of the NaSch model application, the percentiles of travel time distribution are determined on the basis of 500 simulation runs for each queue length. Using the proposed algorithm, the travel time is determined in single simulation run as a fuzzy number $\Theta$ according to the following equation:

$$\theta^{(m)} = \min\{t : x_{l,t}^{(m)} > 333\}, \; m = 1, 2, 3, \tag{15}$$

where $l$ is the index of the last vehicle. The cell number 333 corresponds to location of the stop line at third intersection.

Plots in Fig. 10 show number of vehicles on the modelled road (upstream of the third intersection) for successive time steps of the traffic simulation. Note that the vehicles are inserted into the modelled road only at the beginning of simulation. The number of vehicles decreases during simulation as subsequent vehicles pass over the stop line of the third intersection. The initial number of vehicles is determined by the queue length parameter. The results presented in Fig. 10 a) and b) were obtained for the queue length of 30 and 70 vehicles respectively. The results for NaSch model were estimated after 500 runs of the traffic simulation. In the proposed algorithm, the number of vehicles at time step $t$ is directly calculated for a single simulation run as a fuzzy number $N_t$:

$$n_t^{(m)} = \left|\{i : x_{i,t}^{(m)} \leq 333\}\right|, \; m = 1, 2, 3, \tag{16}$$

where |.| denotes the cardinality of a set.

a) b)

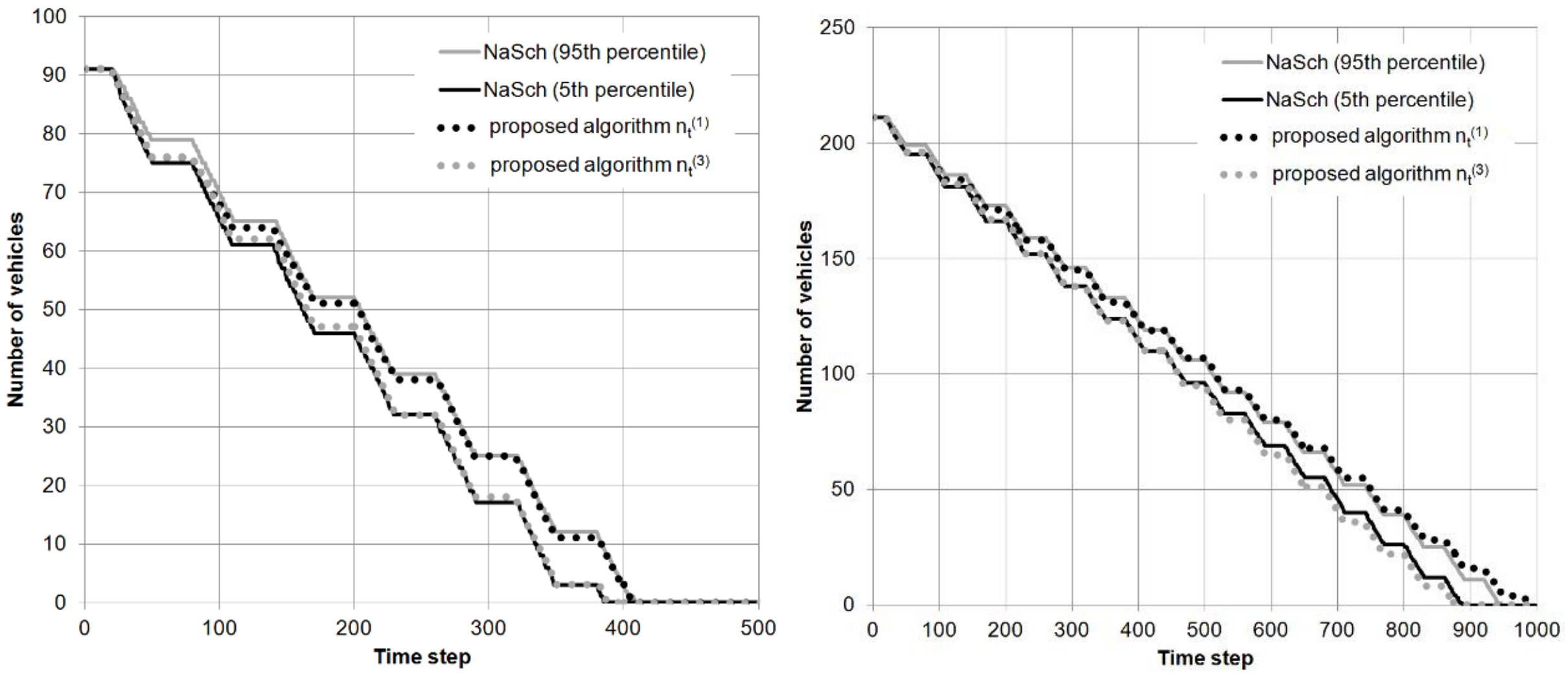

Fig. 10. Number of vehicles on the modelled road for two initial queue lengths: a) 30 b) 70

The results shown in Fig. 9 and Fig. 10 were obtained for the traffic signal cycle of 60 s with green phase of 30 s. The above signal timing parameters were used for all three intersections.

Fig. 11 includes a comparison of flow-density diagrams for the proposed model and the NaSch CA. It is apparent in this plot that the introduced simulation algorithm reflects the spread of traffic flow values ($q$) for the congested branch. Such result is a consequence of the difference between gaps ($g^L = 4$ and $g^H = 3$) produced by the two CA rules (RL, RH) that were used in these experiments. Similar effect for the free-flow branch can be obtained by adopting rules RL and RH with different values of the maximal velocity ($v_{\max}^L < v_{\max}^H$). Note that $v_{\max}^L = v_{\max}^H = 2$ for the experiments reported here.

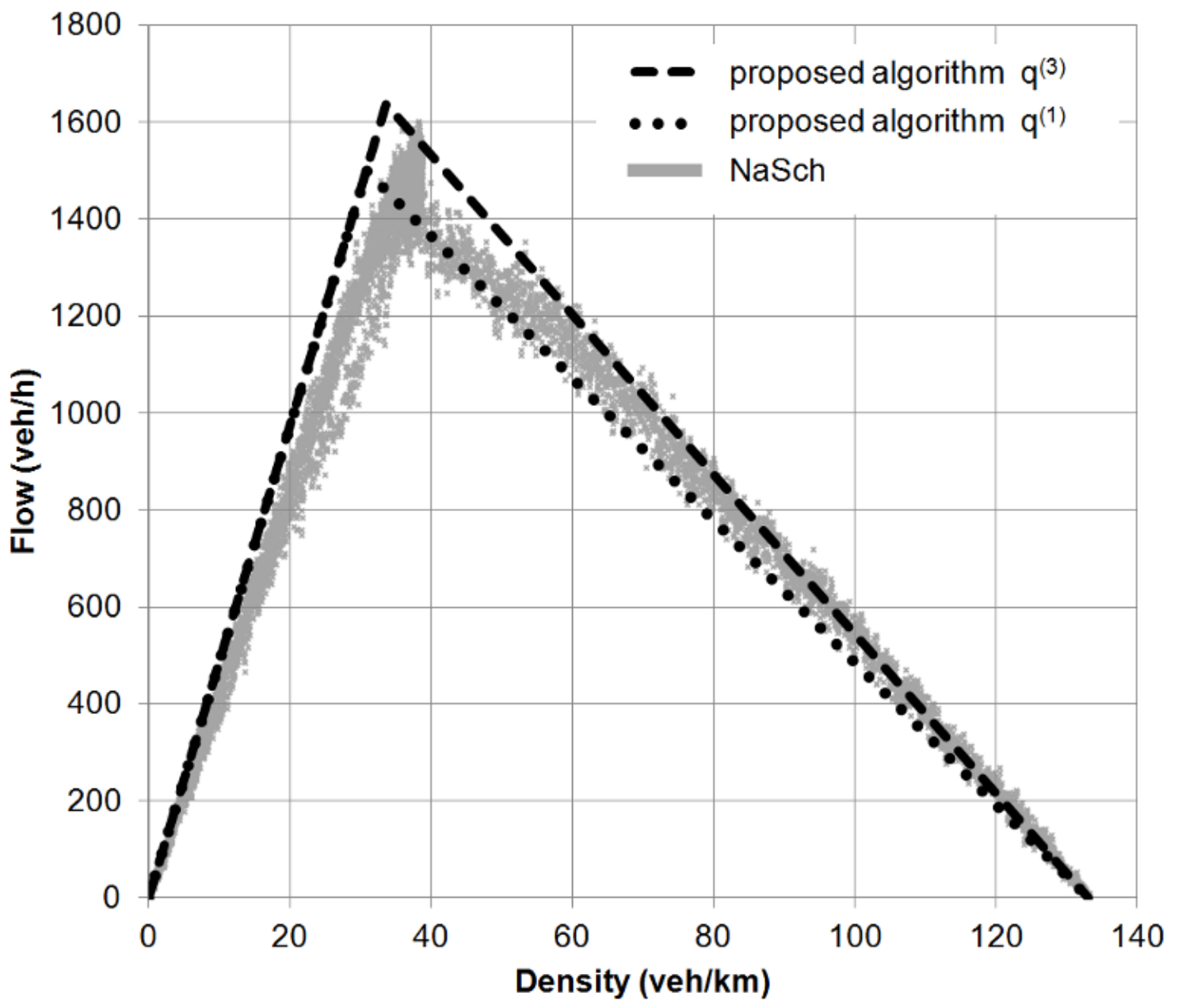

Fig. 11. Flow-density diagrams for the proposed algorithm and the NaSch CA

As it can be observed in the above results, the evolution of the simulated traffic stream is very similar for both algorithms. This fact proves that the proposed algorithm can be appropriately calibrated to reproduce the traffic stream behaviour for a given distribution of saturation flows rates. The experiments have shown that the accuracy of traffic simulation is similar for both considered algorithms. However, the proposed algorithm avoids the aforementioned disadvantages of the existing CA models (Section 3). Firstly, the proposed algorithm can be precisely calibrated by adjusting its parameters. Moreover, uncertainty of the parameters can be taken into account as they are represented by fuzzy numbers. Secondly, the proposed algorithm does not need multiple simulations because it uses the fuzzy numbers to estimate the distributions of traffic performance measures (travel time, the number of vehicles in a given region, delays, queue lengths, etc.) during a single run of the traffic simulation.

## 6 CONCLUSIONS

The CA-based algorithm for simulation of signal controlled traffic streams was proposed by combining CA with fuzzy numbers. The presented approach benefits from advantages of the available CA models and eliminates the main drawbacks that have impeded their applications in traffic control systems. Parameters of the proposed algorithm enable a simple calibration and allow the traffic simulation to reflect predetermined saturation flow rates. The fuzzy numbers are used in order to describe the uncertainty and imprecision of the simulation inputs and outputs. Thus, the imprecise traffic data can be utilised in the proposed approach for the estimation of traffic performance [31]. The experiments reported in this paper show that the traffic simulation with the proposed algorithm is consistent with those performed by stochastic CA. However, the application of the introduced algorithm considerably reduces the computational cost of traffic simulation. These findings are of fundamental importance for real-time applications of CA-based models in the road traffic control.

Although this study focuses on the traffic simulation for signalized streets, the proposed model can be also used for highway traffic. Another potential field of application is the prediction of traffic characteristics and the prediction-based optimization of traffic operations. These directions are open for further research.